\documentclass[twocolumn,showkeys,preprintnumbers,floatfix,amsmath,a4paper,nofootinbib,amssymb]{revtex4}
\newcommand{\beginsupplement}{%
        \setcounter{table}{0}
        \renewcommand{\thetable}{S\arabic{table}}%
        \setcounter{figure}{0}
        \renewcommand{\thefigure}{S\arabic{figure}}%
     }
\usepackage{graphicx}
\usepackage{float}
\usepackage[T1,T2A]{fontenc}
\usepackage[utf8x]{inputenc}
\usepackage[russian,english]{babel}
\usepackage{textcomp}

\begin{document}
\selectlanguage{english}

\title{Schools are segregated by educational outcomes in the digital space}

\author{Ivan Smirnov$^1$}

\thanks{ibsmirnov@hse.ru}

\affiliation{
$^1$ Institute of Education; National Research University Higher School of Economics, Myasnitskaya ul., 20, Moscow 101000, Russia}

\begin{abstract}
The Internet provides students with a unique opportunity to connect and maintain social ties with peers from other schools, irrespective of how far they are from each other. However, little is known about the real structure of such online relationships. In this paper, we investigate the structure of interschool friendship on a popular social networking site. We use data from $36,951$ students from $590$ schools of a large European city. We find that the probability of a friendship tie between students from neighboring schools is high and that it decreases with the distance between schools following the power law. We also find that students are more likely to be connected if the educational outcomes of their schools are similar. We show that this fact is not a consequence of residential segregation. While high- and low-performing schools are evenly distributed across the city, this is not the case for the digital space, where schools turn out to be segregated by educational outcomes. There is no significant correlation between the educational outcomes of a school and its geographical neighbors; however, there is a strong correlation between the educational outcomes of a school and its digital neighbors. These results challenge the common assumption that the Internet is a borderless space, and may have important implications for the understanding of educational inequality in the digital age.
\end{abstract}

\maketitle
Even Pope said so \cite{pope2014}. The Internet creates unique opportunities for people to connect with each other. It may, therefore, be significantly beneficial for its users because social ties are known to play a significant role in human well-being including life-satisfaction \cite{diener1999subjective}, health \cite{holt2010social,kawachi2001social}, and professional development \cite{podolny1997resources,ng2005predictors}. There is growing evidence that these findings apply not only to offline social ties but to online friendship as well \cite{hobbs2016online,manago2012me}. This role of the internet may be particularly important for underprivileged groups of people such as students from low-performing schools who lack resources in their immediate environment. Connections with students from high-performing schools might potentially influence their university aspirations \cite{cohen1983peer}, improve educational outcomes \cite{lomi2011some}, and promote positive behavioral change \cite{maxwell2002friends}.

People from underprivileged backgrounds tend not to benefit as much as their peers from the Internet (a phenomenon usually referred to as digital inequality \cite{dimaggio2004unequal}). While well-educated people often use the Internet for medical or juridical advice, job seeking or education, their less educated peers use it predominantly for entertainment \cite{pearce2017somewhat,buchi2016modeling,van2014digital}. The use of social media by students is known to be differentiated in a similar way depending on their academic performance. High-performing students use it for information seeking while low-performing students for chatting and entertainment \cite{junco2012too,smirnov2018predicting}. It may be expected that online social ties would also depend on academic achievements and that students might be segregated by the educational outcomes in the digital space. At a general level, segregation is the degree to which several groups of people are separated from each other \cite{allen2007should}. In this paper, we investigate whether students from high- and low-performing schools are separated (i.e. not connected via online friendship) in the digital space.

\begin{figure}[!ht]
\centering
\includegraphics[width=1.0\linewidth]{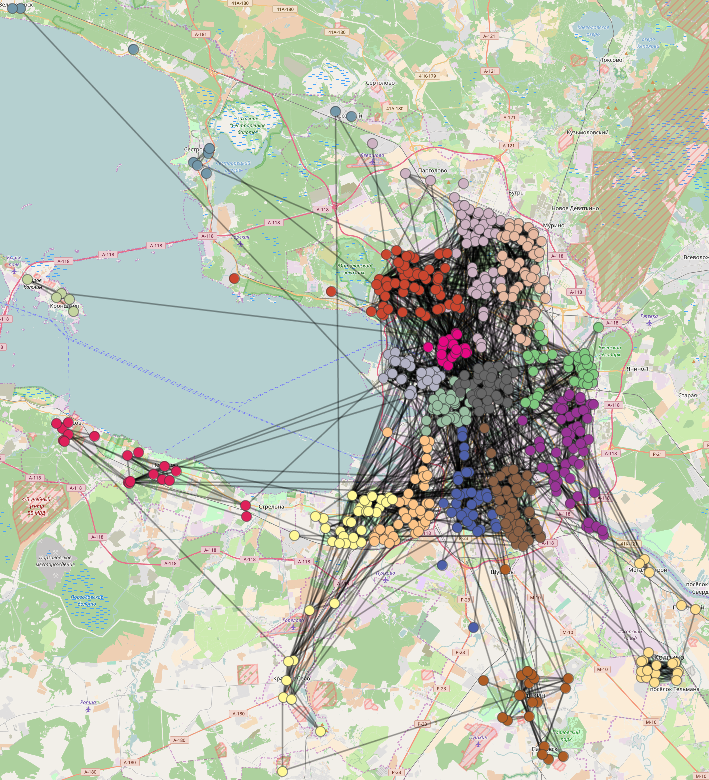}
\caption{\textbf{The school network.} Circles represent schools. Different colors correspond to administrative districts of Saint Petersburg. Two schools are connected if there is a friendship tie between their students. For visual clarity, only strong connections (at least three friendship ties) are shown.}
\label{fig:Network}
\end{figure}
\newpage
We use data from $36,951$ 15-year-old students from $590$ schools of Saint Petersburg, Russia, registered on a popular social networking site VK\footnote{http://vk.com} (see Methods for details about the sample). VK is the Russian analog of Facebook and the largest European social networking site. It is ubiquitous among young Russians: more than 90\% of 18-24-year-olds use it regularly \cite{fom2016online}. The information in users’ public profiles includes their age and the schools they are studying in. This information is available via the open application programming interface (API) of VK. We use the VK API to download information about all students who indicate that they study in one of Saint Petersburg’s schools and who were born in 2001 (i.e. that students were 15 years old at the time of data collection).

Similar to other social networking sites, users might become “friends” on VK if they mutually confirm this status. We use information about such online friendships to construct a weighted network of schools (Fig. \ref{fig:Network}), where two schools are connected if there is at least one friendship tie between their students (see Methods for details), and the weight corresponds to the number of such ties. For each school, the information about its geographical coordinates along with the performance of its graduates on the unified state examination (USE) is available (see Methods). The USE scores serve as a proxy for schools’ educational outcomes.

Residential segregation by income is believed to be an important source of variation in schools’ educational outcomes in some countries \cite{flores2008residential,gordon2003urban,owens2018income}. It means that low-performing schools are concentrated in less affluent neighborhoods and the educational outcomes of a school could be effectively predicted from the socioeconomic status of its district \cite{reardon2017geography}. The situation might be different in Saint Petersburg thanks to the egalitarian nature of the Russian educational system inherited from the Soviet period. To account for potential effects of residential segregation, we collect data from $11,034$ apartments from the largest Russian real estate site CIAN\footnote{http://cian.ru} and use average apartment price as a proxy of neighborhood affluence. We then check whether schools’ educational outcomes are correlated with the affluence of their neighborhood.

We measure geographical segregation of schools as a correlation between the educational outcomes of a school and those of its closest geographical neighbors. We then compare this segregation with that in the digital space. In this case, instead of the closest geographical neighbors, we examine the educational outcomes of schools’ closest digital neighbors. We assume that the distance between two schools in the digital space is inversely proportional to the number of online friendship ties between them.

The probability of an online friendship between two people is known to be strongly dependent on the geographical distance between them \cite{takhteyev2012geography,shin2015new,lengyel2015geographies,grabowicz2014entangling}. It is, therefore, important to ensure that any observed effect for the digital network of schools is not solely driven by the geographical constraints. To achieve this, we use a random graph model that preserves geographical constraints – namely, the probability of a friendship tie between two schools given the geographical distance between them. We then compare the results obtained for such random networks with the observed results for the real network.

\begin{figure}
\centering
\includegraphics[width=1.0\linewidth]{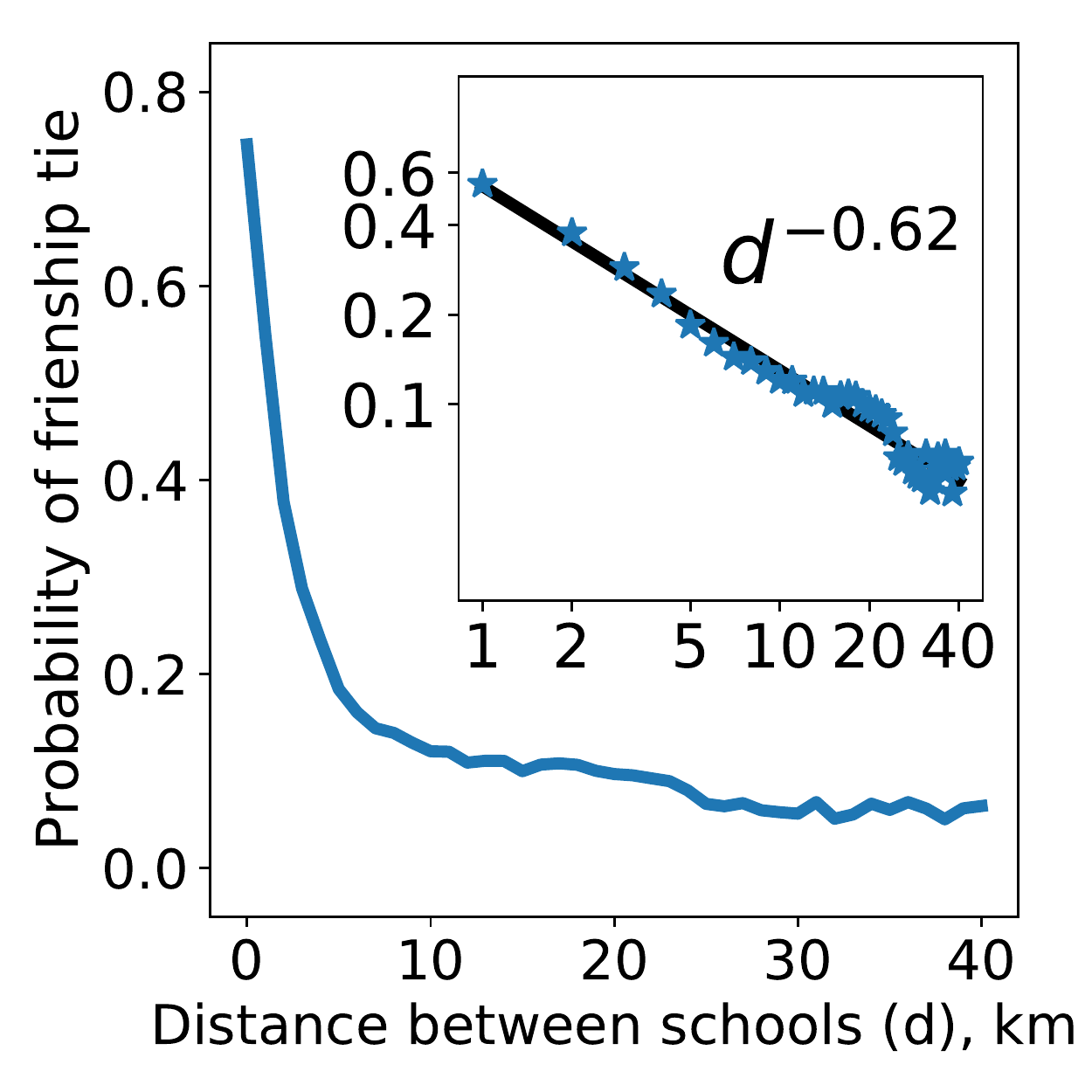}
\caption{\textbf{Probability of a friendship tie between two schools as a function of distance between these schools.} For close schools, the probability is $0.75$ and it then declines with distance following the power law (inset).}
\label{fig:Distance}
\end{figure}

\section*{Results}
\textbf{Distance and online relationships}\newline
We find that geographical distance plays an important role in the formation of an interschool friendship. The probability of a friendship tie between two close schools is high ($0.75$) but it declines rapidly with distance following the power law (Fig. \ref{fig:Distance}). The best fit is provided by the exponent $-0.62$ (Fig. \ref{fig:Distance} inset), which is similar to the previously observed results \cite{grabowicz2014entangling}.\newline

\begin{figure*}[!ht]
\centering
\includegraphics[width=0.48\linewidth]{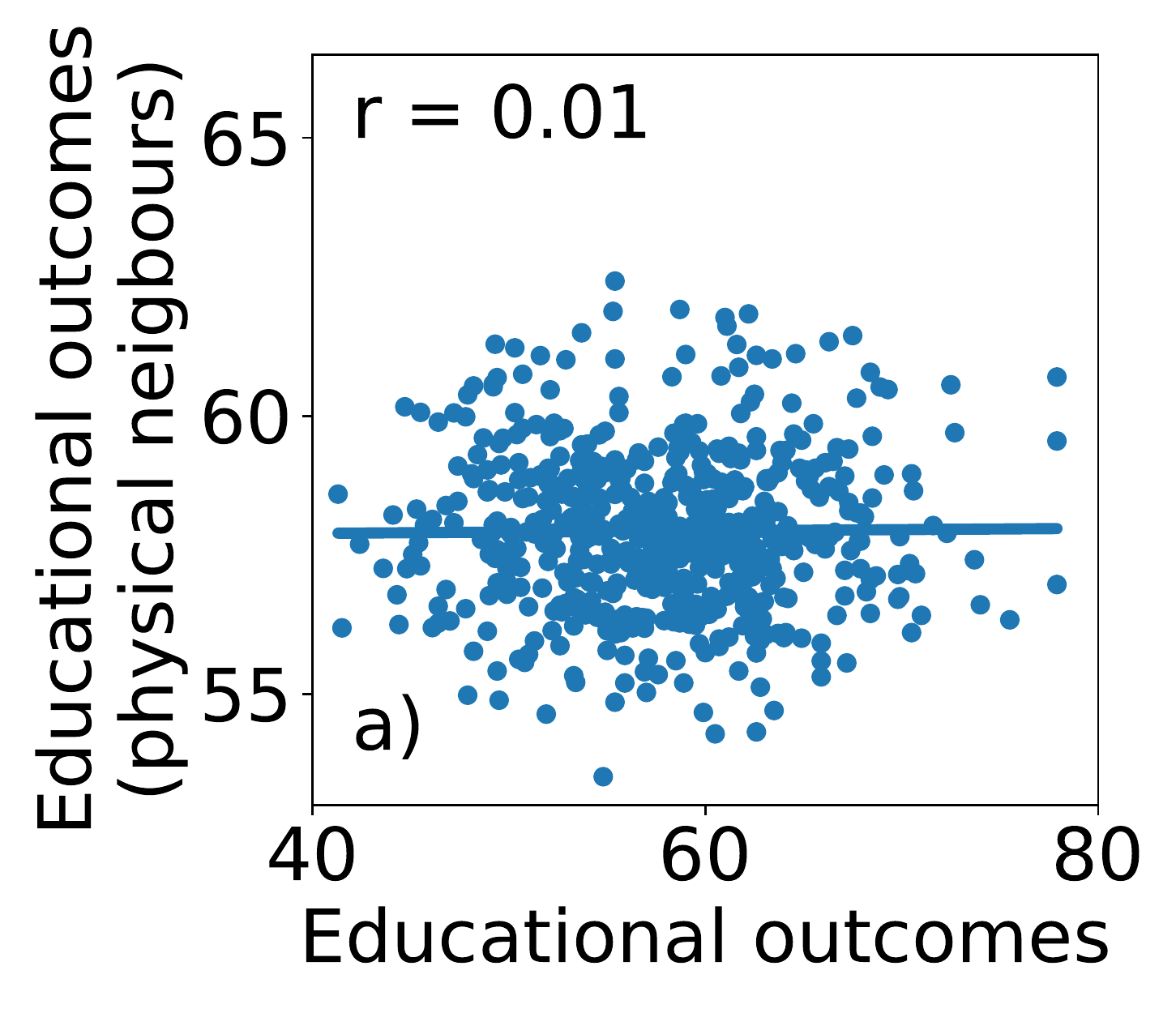}
\includegraphics[width=0.48\linewidth]{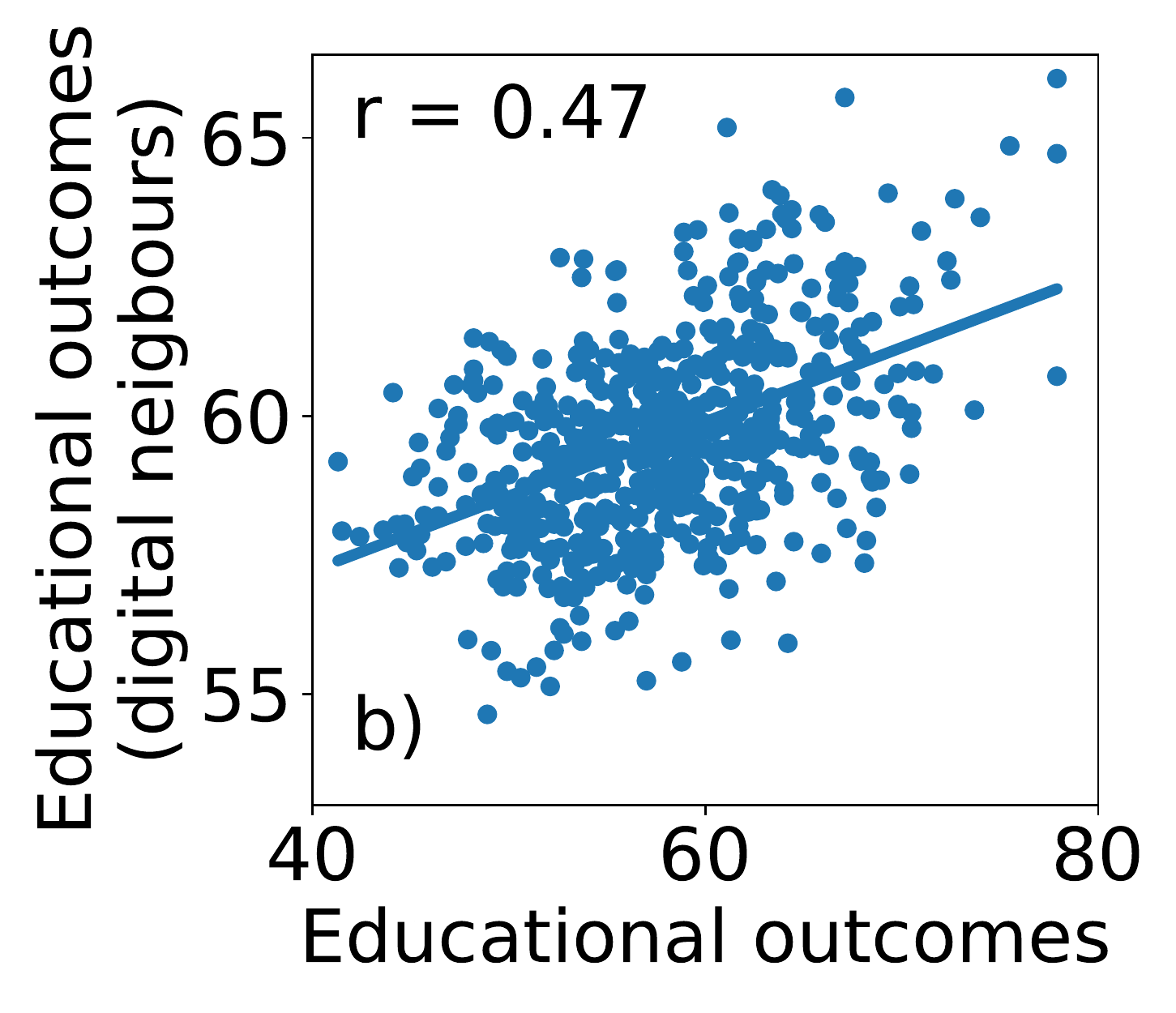}
\caption{\textbf{Correlation between educational outcomes of schools and their $20$ closest geographical (a) and digital neighbors (b).} While there is no correlation for physical neighbors, there is a relatively strong correlation for digital neighbors. These results hold true regardless of the number of neighbors used in the analysis.}
\label{fig:Segregation}
\end{figure*}

\begin{figure}[H]
\centering
\includegraphics[width=1.0\linewidth]{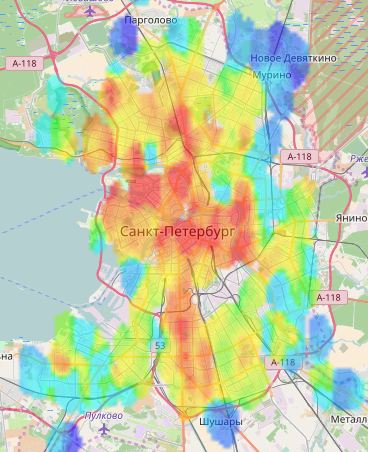}
\caption{\textbf{The heat map of average apartment price in Saint Petersburg.} The price is the highest in the city center, near Krestovsky Island, and along the Moscow avenue.
}
\label{fig:Heat}
\end{figure}

\begin{figure}[H]
\centering
\includegraphics[width=1.0\linewidth]{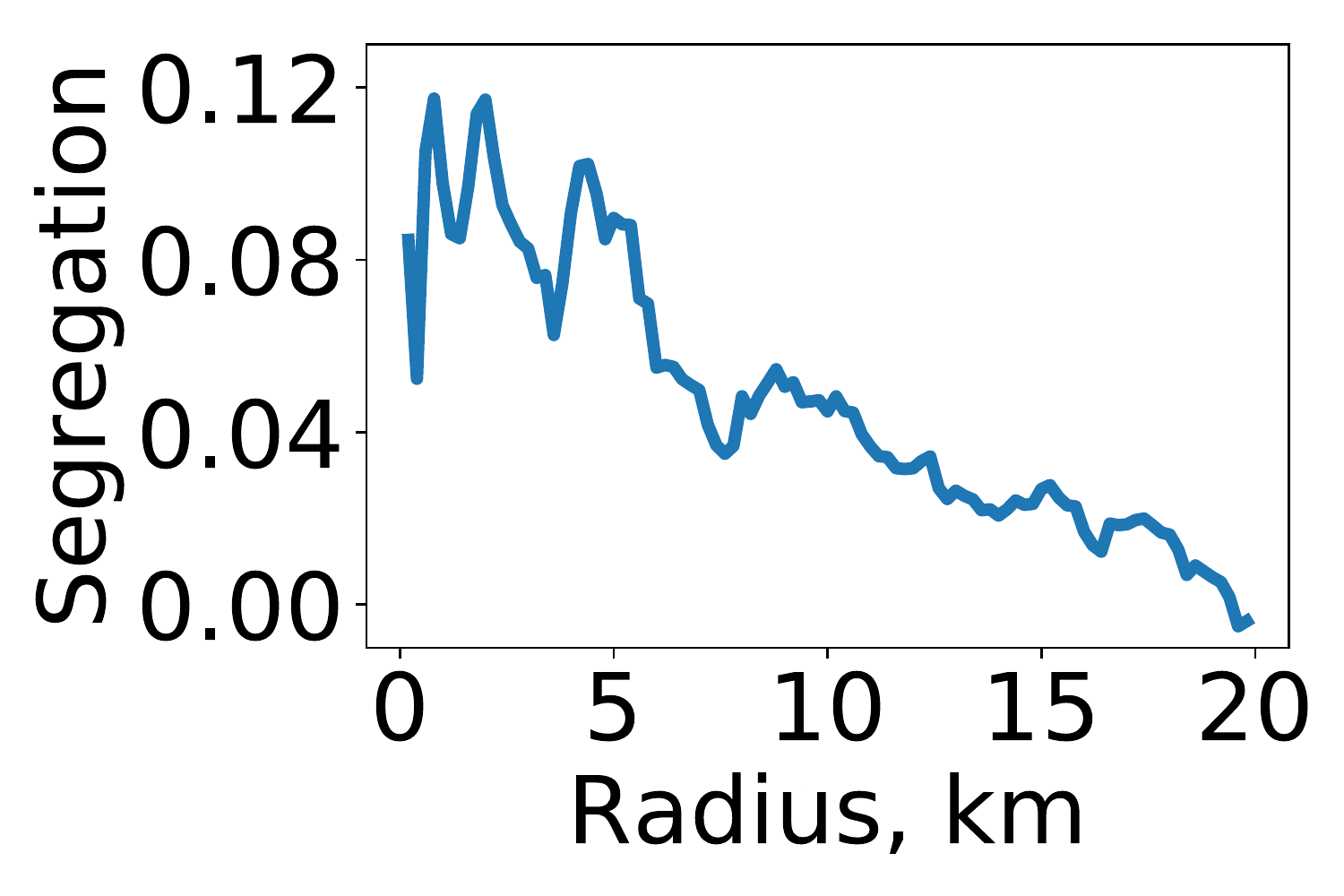}
\caption{\textbf{Segregation $S_n(R)$ as a function of the radius R that defines school neighborhood.}}
\label{fig:Radius}
\end{figure}

\begin{figure}[H]
\centering
\includegraphics[width=1.0\linewidth]{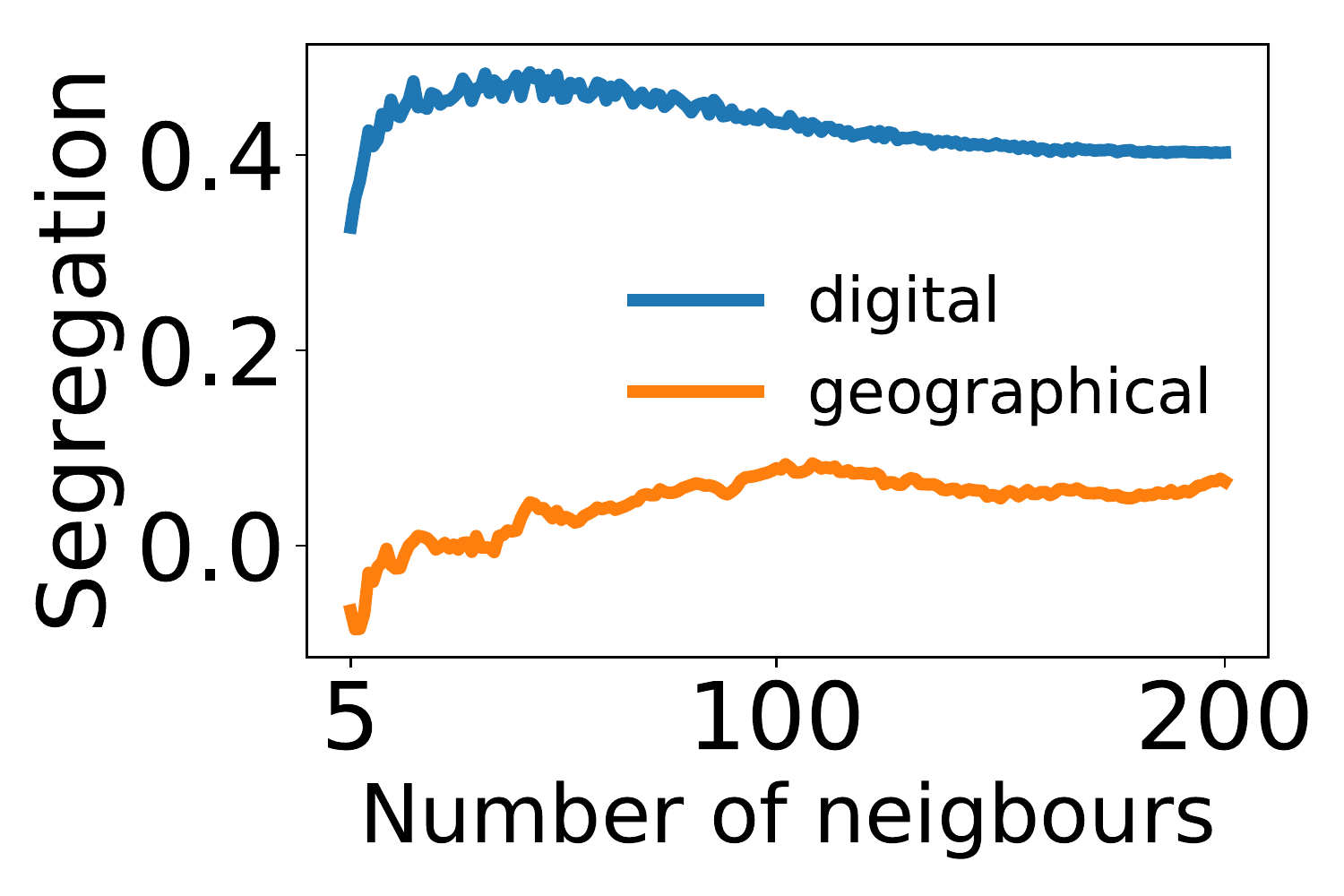}
\caption{\textbf{Digital $S_d(N)$ and geographical $S_g(N)$ segregations as functions of the number of the neighbors N used in the analysis.}}
\label{fig:Neibgours}
\end{figure}

\textbf{Geographical segregation}\newline
We find that the educational outcomes of schools do not depend on their distance from the city center (Pearson correlation coefficient between USE scores of schools and their distance from the center is $0.018$, $P = 0.65$). The distance from the center may be, however, a poor proxy for neighborhood affluence. Hence, we additionally collect information about average apartment prices across the city (see Fig. \ref{fig:Heat} for the corresponding heat map). We then compute the correlation between schools’ USE scores and neighborhood affluence, $S_n(R)$ (see Methods). The exact value depends on $R$ (see Fig. \ref{fig:Radius}), and the maximum value is $S_n = 0.12$ $(P = 0.007)$, indicating a weak correlation between educational outcomes and neighborhood affluence. Finally, we compute a correlation between USE scores of schools and average USE score of their N closest geographical neighbors, $S_g(N)$ (see Methods). We find no correlation $S_g(N) = 0.01$ $(P = 0.73)$ for $N = 20$ (Fig. \ref{fig:Segregation}a); this result holds true for all values of $N$ (Fig. \ref{fig:Neibgours}).

We therefore find that there is only a weak if any relationship between educational outcomes of a school and its location in physical space. However, as we show in the next section, this result does not apply for the school location in the digital space.\newline

\textbf{Digital segregation}\newline
We find that there is a relatively strong correlation between the educational outcomes of schools and their N closest digital neighbors (see Methods). $S_d(N) = 0.47$ $(P < 10^{-33})$ for $N = 20$ (Fig \ref{fig:Segregation}b). The correlation is significant for all $N$ (Fig. \ref{fig:Neibgours}). 

To rule out the role of geographical constraints in the observed digital segregation, we use a random graph model that preserves relationships between distance and probability of a friendship tie from the observed network (i.e. we create a tie between two schools with a probability from distribution represented in Figure \ref{fig:Distance} that depends on distance between schools). We compute $S^{rand}_d(1)$ for generated random networks and compare it with $S_d(1)$. After $10,000$ simulations we obtain $\langle S^{rand}_d(1) \rangle = -0.0005$ and $SD(S^{rand}_d(1)) = 0.04$. The maximum value $\max(S^{rand}_d(1)) =  0.14 < S_d(1)$. This result makes the observed digital segregation significant with $P < 10^{-4}$.

We also find that high-performing schools not only tend to be connected with each other but also have more connections on average than low-performing schools. The correlation between the degree centrality of schools in the network and their educational outcomes is $0.49$. Note that this simple network property explains as much variation in the educational outcomes as the socioeconomic status of students \cite{yasterbov2014contextualizing}, which is one of the key context variables used in educational studies \cite{sirin2005socioeconomic}. 

We show, therefore, that the educational outcomes of a school are closely related to its location in the digital space. More central schools tend to be high performing. We also show that schools with similar academic performance tend to be connected in the digital space. We demonstrate that these results cannot be explained by schools’ locations in the physical space.

\section*{Discussion}
Both for research and policy-making purposes, it is crucial to understand the context in which schools operate. This requirement traditionally means collecting information about school resources and the socioeconomic status of its students. Today, students spend much of their time online \cite{koroleva2016always}, and it may be warranted to consider students’ online environment on a par with their home environment. In this paper, we focus only on one dimension of such an online environment, namely interschool friendship on a social networking site. We find that school position in an online friendship network could explain as much variation in the educational outcomes of its students as their socioeconomic status, indicating the importance of the digital context. Online inequalities might merely reflect existing socioeconomic inequality or rather complement it. Future research is required to clarify this relationship. 

Social media have become the main source of information for young people. In Russia, VK is referred to as the main source of information about the country and the world by 70.3\% of respondents — more than any other information source \cite{kasamara2017}. It is also considered more trustworthy than traditional media \cite{kasamara2017}. The news feed of the social network mainly comprises posts shared by online friends. The friends from different schools may, therefore, be an important source of diversity in the information environment of students. In particular, the connections with students from high-performing schools could have a positive impact on students from low-performing schools. However, our results suggest that interschool friendship ties mainly exist between schools with similar educational outcomes. Intriguingly, this digital separation cannot be explained by the geographical location of schools. This result means that the digital environment not only fails to remove segregation, but rather might amplify it.

\section*{Methods}
\textbf{Data collection}\newline
According to the open data government portal\footnote{http://data.gov.spb.ru}, there are 638 high schools in Saint Petersburg. This number excludes specific types of schools such as boarding schools, cadet schools, and educational centers. We use open VK API to find these schools in the VK database. We find VK IDs for 628 of the schools. We exclude school \textnumero1 from the sample because it has an unreasonable number of users (more than 1000 per cohort). We also exclude two pairs of schools with identical names. We then use data from the web portal "Schools of Saint Petersburg"\footnote{http://www.shkola-spb.ru} to obtain the average performance of schools’ graduates at the Unified State Examination. This is a mandatory state examination that all school graduates should pass in Russia. This information was available for 590 schools from our sample.

We then perform requests to VK API to obtain the lists of all users who were born in 2001 and indicate that they are studying in one of the schools from our sample. To exclude users who provided false information about their school, we remove profiles with no friends from the same school, as previously recommended \cite{smirnov2016search}. We also exclude students who indicate several schools in their profiles. Finally, we download the lists of all VK friends for users from our sample. All collected data is publicly available. The VK team confirmed to us that we can use its API in this way for research purposes.

We also use data from the largest Russian real estate site CIAN to collect information about the prices of all 2-room apartments in Saint Petersburg listed on the site. For each apartment, its price per square meter was calculated. CIAN team approved the use of this data for research purposes.\newline 

\textbf{Network of schools}\newline
We define a $36,951\times 36,951$ adjacency matrix $F$ that represents the friendship network of students (i.e. $F_{i,j} = 1$ if students $i$ and $j$ are friends on VK and $F_{i,j} = 0$ otherwise). We assume that student $i$ studies in school $s(i)$, and construct a weighted network of schools by counting the number of all friendship ties between two schools. This network is represented by $590\times 590$ matrix $A$ where $$A_{k,l} = \sum\limits_{\{i,j | s(i) = k, s(j) = l\}} F_{i,j}.$$ One potential disadvantage of this definition is that two schools could be considered as closely connected when only one student from the first school has a lot of friends from the other. We therefore also use an alternative way to define the weight of the school tie. In this case, instead of friendship ties, we count the number of students from one school that have friends from another (i.e. we define $\tilde{A}_{k,l} = |\{i | s(i) = k~\text{and}~\exists j: F_{i,j} = 1, s(j) = l\}|$). We could then construct a symmetric matrix $\hat{A}_{k,l} = \min(\tilde{A}_{k,l}, \tilde{A}_{l,k})$. This alternative metric leads to the same results, and therefore we opted for the first more straightforward approach.\newline

\textbf{Segregation measures}\newline
If $U_i$ is the average performance on the Unified State Examination of graduates from school $i$, we could then define segregation based on the affluence of school neighborhoods in the following way:
$$S_n(R) = \text{corr}(U_i, \sum\limits_{\{j | d(i, j) < R\}} P_j / |\{j | d(i, j) < R\}|),$$

where $P_j$ is the price of apartment $j$ in rubles per square meter and $d(i, j)$ is the distance between school $i$ and apartment $j$.

We denote geographical neighbors of school $i$ by $N_g(i)$. $N_g(i) =  (s_{i,1}, ..., s_{i,590})$ is an ordered list of all schools such as $\tilde{d}(i, s_{i,k}) <= \tilde{d}(i, s_{i,k+1})$, where $\tilde{d}$ is the geographical distance between schools. We then denote the list of $k$-closest geographical neighbors by $N_g^k(i) = (s_{i,1}, ..., s_{i,k})$. We define the $k$-closest digital neighbors $N_d^k(i)$ by replacing geographical distance with the digital distance that is equal to $1/A_{i,j}$.

We then define geographical and digital segregations by academic performance in the following manner:  $$S_g(k) = \text{corr}(U_i, \sum\limits_{\{j | j \in N_g^k(i)\}} U_j / k)$$
$$S_d(k) = \text{corr}(U_i, \sum\limits_{\{j | j \in N_d^k(i)\}} U_j / k)$$

Note that in the case of digital segregation, there could be several schools with exactly the same distance from a certain school. In this case, $N_d^k(i)$ is not uniquely defined. In our computations, we randomly select with equal probabilities one of the possible $N_d^k(i)$.

\clearpage
\bibliographystyle{unsrt}
\bibliography{schoolsegregation}
\clearpage
\beginsupplement

\end{document}